\documentclass[12pt]{article}	
\usepackage{amsmath,amssymb,bm,graphicx}
\usepackage{here}
\usepackage{color} 	

\begin{document}

\begin{flushright}
%July, 2019
\end{flushright}

\vskip 0.5 truecm

\begin{center}
{\Large{\bf  Lensing of Dirac monopole in Berry's phase 
}}
\end{center}
\vskip .5 truecm
\centerline{\bf Kazuo Fujikawa~$^1$
 {\rm and} Koichiro Umetsu~$^2$ }
\vskip .4 truecm
\centerline {\it $^1$~Interdisciplinary Theoretical and Mathematical Sciences Program, 
}
\centerline {\it   RIKEN, Wako 351-0198, 
Japan}
\vskip 0.4 truecm
\centerline {\it $^2$~Laboratory of Physics, College of Science and Technology, and Junior College, }
\centerline{\it Funabashi Campus, Nihon University, Funabashi, Chiba 274-8501, Japan}
\vskip 0.5 truecm

\makeatletter
%\@addtoreset{equation}{section}
%\def\theequation{\thesection.\arabic{equation}}
\makeatother
\begin{abstract}
Berry's phase, which is associated with the slow cyclic motion with a finite period, looks like a Dirac monopole when seen from far away but smoothly changes to a dipole near the level crossing point in the parameter space in an exactly solvable model. 
This topology change of Berry's phase is visualized as a result of lensing effect; the monopole supposed to be located at the level crossing point appears at the displaced point when the variables of the model deviate from the precisely adiabatic movement. The effective magnetic field generated by Berry's phase is determined by a simple geometrical consideration of the magnetic flux coming from the displaced Dirac monopole.

\end{abstract}

%\large

\section{Monopole in Berry's phase}
The notion of topology and topological phenomena  have become common in various fields in physics. Among them, topological Berry's phase arises when one analyzes level crossing phenomena in quantum mechanics by a careful use of the adiabatic theorem \cite{Higgins, Berry, Simon}. The basic mechanism of the phenomenon is very simple and it is ubiquitous in quantum physics. It is thus surprising that one encounters Dirac's magnetic monopole-like topological phase \cite{Dirac}  essentially at each level crossing point for the sufficiently slow cyclic motion in quantum mechanics \cite{Berry, Fujikawa1}.  The general aspects of  the monopole-like topological Berry's phase in the adiabatic limit  and the smooth change of Berry's phase to a dipole in the nonadiabatic limit have been analyzed in \cite{DF1} using an exactly solvable version of Berry's model \cite{Fujikawa1}. We here report on a more quantitative description of the magnetic field generated by Berry's phase, which is essential to understand the motion of a   particle placed in the monopole-like field, together with a surprising connection of the topology change of Berry's phase  with the formal geometrical movement of Dirac's monopole  in the parameter space caused by the nonadiabatic variation of parameters. This movement is characterized as an analogue of the lensing effect of Dirac's monopole in Berry's phase.

 We first briefly summarize the essential setup of the problem for the sake of completeness. 
Berry originally analyzed the Schr\"{o}dinger equation \cite{Berry}
\begin{eqnarray}\label{starting equation}
i\hbar\partial_{t}\psi(t)=\hat{H}\psi(t)
\end{eqnarray}
for the Hamiltonian
$\hat{H}=-\mu\hbar\vec{\sigma}\cdot\vec{B}(t)$
describing the motion of a magnetic moment $\mu\hbar\vec{\sigma}$ placed in a rotating magnetic field 
\begin{eqnarray}\label{solvable model}
\vec{B}(t)=B(\sin\theta\cos\varphi(t), 
\sin\theta\sin\varphi(t),\cos\theta )
\end{eqnarray}
with $\vec{\sigma}$ standing for Pauli matrices. The level crossing takes place at the vanishing external field $B=0$. 
It is explained later that this parameterization \eqref{solvable model} describes the essence of Berry's phase. 
It has been noted that the equation \eqref{starting equation} is exactly solved if one restricts the movement of the magnetic field to the form
$\varphi(t)=\omega t$ with constant $\omega$, and constant $B$ and $\theta$ \cite{Fujikawa1}.
The exact solution is then written as
 \begin{eqnarray}\label{eq-exactamplitude1}
\psi_{\pm}(t)
&=&w_{\pm}(t)\exp\left[-\frac{i}{\hbar}\int_{0}^{t}dt
w_{\pm}^{\dagger}(t)\big(\hat{H}
-i\hbar\partial_{t}\big)w_{\pm}(t)\right]\nonumber\\
&=&w_{\pm}(t)\exp\left[-\frac{i}{\hbar}\int_{0}^{t}dt
w_{\pm}^{\dagger}(t)\hat{H}w_{\pm}(t)\right]\exp\left[-\frac{i}{\hbar}\int_{0}^{t}
{\cal A}_{\pm}(\vec{B})\cdot\frac{d\vec{B}}{d t}dt\right]
\end{eqnarray}
where 
\begin{eqnarray}\label{exact eigenfuntion}
w_{+}(t)&=&\left(\begin{array}{c}
            \cos\frac{1}{2}(\theta-\alpha) e^{-i\varphi(t)}\\
            \sin\frac{1}{2}(\theta-\alpha)
            \end{array}\right), \ \ \ 
w_{-}(t)=\left(\begin{array}{c}
            \sin\frac{1}{2}(\theta-\alpha) e^{-i\varphi(t)}\\
            -\cos\frac{1}{2}(\theta-\alpha)
            \end{array}\right).
\end{eqnarray}
It is important that these solutions differ from the so-called instantaneous solutions used in the adiabatic approximation, which are given by  setting $\alpha=0$; the following analysis of topology change is not feasible using the instantaneous solutions.   The parameter $\alpha(\theta,\eta)$ is defined by $\mu\hbar B\sin\alpha= (\hbar\omega/2)\sin(\theta-\alpha)$ or equivalently \cite{Fujikawa1}
\begin{eqnarray}\label{cotangent}
\cot\alpha(\theta,\eta)=\frac{\eta+\cos\theta}{\sin\theta}
\end{eqnarray}
with $\eta=2\mu\hbar B/\hbar\omega$ for $0\leq \theta\leq\pi$, which specifies the branch of the cotangent function.  
The second term in the exponential of the exact solution \eqref{eq-exactamplitude1}
is customarily called Berry's phase which is defined by a potential-like object (or connection)
\begin{eqnarray}
{\cal A}_{\pm}(\vec{B})\equiv w_{\pm}^{\dagger}(t)(-i\hbar\frac{\partial}{\partial \vec{B}})w_{\pm}(t).
\end{eqnarray}
This potential describes an azimuthally symmetric static magnetic monopole-like object in the present case.

The solution \eqref{eq-exactamplitude1} is confirmed by evaluating
\begin{eqnarray}
i\hbar\partial_{t}\psi_{\pm}(t)
&=&\{ i\hbar\partial_{t}w_{\pm}(t)+w_{\pm}(t)[w_{\pm}^{\dagger}(t)\big(\hat{H}
-i\hbar\partial_{t}\big)w_{\pm}(t)]\}\nonumber\\
&&\times\exp\left[-\frac{i}{\hbar}\int_{0}^{t}dt^{\prime}
w_{\pm}^{\dagger}(t^{\prime})\big(\hat{H}
-i\hbar\partial_{t^{\prime}}\big)w_{\pm}(t^{\prime})\right]\nonumber\\
&=&\{ i\hbar\partial_{t}w_{\pm}(t)+w_{\pm}(t)[w_{\pm}^{\dagger}(t)\big(\hat{H}
-i\hbar\partial_{t}\big)w_{\pm}(t)]\nonumber\\
&&\   +w_{\mp}(t)[w_{\mp}^{\dagger}(t)\big(\hat{H}
-i\hbar\partial_{t}\big)w_{\pm}(t)]\}\nonumber\\
&&\times\exp\left[-\frac{i}{\hbar}\int_{0}^{t}dt^{\prime}
w_{\pm}^{\dagger}(t^{\prime})\big(\hat{H}
-i\hbar\partial_{t^{\prime}}\big)w_{\pm}(t^{\prime})\right]\nonumber\\
&=&\hat{H}\psi_{\pm}(t)
\end{eqnarray}  
where we used
$w_{\mp}^{\dagger}\big(\hat{H}-i\hbar\partial_{t}\big)w_{\pm}=0$ by noting \eqref{cotangent}, and the completeness relation $w_{+}w_{+}^{\dagger}+w_{-}w_{-}^{\dagger}=1$. 

The parameter 
$\eta\geq 0$ is written as 
\begin{eqnarray}\label{parameter2}
\eta=\frac{2\mu\hbar B}{\hbar\omega}=\frac{\mu BT}{\pi}
\end{eqnarray}
when one defines the period $T=2\pi/\omega$. The parameter $\eta$ is a ratio of the two different energy scales appearing in the model, namely, the static energy $2\mu\hbar B$ of the dipole moment in an external magnetic field  and the kinetic energy (rotation energy) $\hbar\omega$: $\eta\gg 1$  (for example, $T\rightarrow \infty$ for any finite $B$) corresponds to the adiabatic limit, and $\eta\ll 1$ (for example, $T\rightarrow 0$ for finite $B$) corresponds to the nonadiabatic limit. In a mathematical treatment of the adiabatic theorem, the precise adiabaticity  is defined by $T\rightarrow \infty$ with fixed $B$ \cite{Simon}. 

The parameter $\alpha(\theta,\eta)$ in \eqref{cotangent} is normalized as $\alpha(0,\eta)=0$
by definition. Then the topology of the monopole-like object is specified by the value 
\begin{eqnarray}
\lim_{\theta\rightarrow\pi} \alpha(\theta, \eta) = 0, \frac{1}{2}\pi, \pi,
\end{eqnarray}
for $\eta>1$, $\eta=1$ and $\eta<1$, respectively,
as is explained later.

The extra phase factor for one period of motion is written as,
\begin{eqnarray}\label{Berry's phase1}
\exp\left[-\frac{i}{\hbar}\oint
{\cal A}_{\pm}(\vec{B})\cdot\frac{d\vec{B}}{d t}dt\right]&=&\exp\{-i\oint \frac{-1\mp\cos(\theta-\alpha(\theta,\eta))}{2}d\varphi \}\nonumber\\
&=&\exp\{-i\oint\frac{1\mp\cos(\theta-\alpha(\theta,\eta))}{2}d\varphi+2i\pi \}\nonumber\\
&=&\exp\{-\frac{i}{\hbar}\Omega_{\pm} \},
\end{eqnarray}
with the monopole-like integrated flux
\begin{eqnarray}\label{solid-angle}
\Omega_{\pm} 
&=&\hbar\oint \frac{[1\mp\cos(\theta-\alpha(\theta,\eta))]}{2B\sin\theta}B\sin\theta d\varphi .
\end{eqnarray}
In \eqref{Berry's phase1}, we adjusted the trivial phase $2\pi i$ for the convenience of the later analysis; this is related to a  gauge transformation of Wu and Yang~\cite{Wu-Yang, DF1}.  
The corresponding energy eigenvalues are
\begin{eqnarray}\label{energy eigenvalue}
E_{\pm}= w_{\pm}^{\dagger}(t)\hat{H}w_{\pm}(t) =\mp (\mu\hbar B\cos\alpha).
\end{eqnarray}
 From now on, we concentrate on $\Omega_{+}$ associated with the energy eigenvalue $E_{+}$; the monopole $\Omega_{-}$ associated with the nergy eigenvalue $E_{-}$ is described by $-\Omega_{+}$ up to a gauge transformation of Wu and Yang.
 We then have an {\em azimuthally symmetric}  monopole-like potential \cite{DF1}
\begin{eqnarray}\label{new potential}
{\cal A}_{\varphi} 
= \frac{\hbar }{2B\sin\theta} [1 - \cos\Theta(\theta, \eta) ]
\end{eqnarray}
and ${\cal A}_{\theta} ={\cal A}_{B} =0$, where we defined 
\begin{eqnarray}\label{effective angle}
\Theta(\theta, \eta)=\theta-\alpha(\theta,\eta).
\end{eqnarray}
The standard Dirac monopole \cite{Dirac} is recovered when one sets $\alpha(\theta,\eta)=0$ (or in the ideal adiabatic limit $\eta=\infty$ in \eqref{cotangent}), namely, $\Theta=\theta$ in \eqref{new potential} and when $B$ is identified with the radial coordinate $r$ in the real space.
The crucial parameter $\Theta(\theta, \eta)$ is shown in Fig.1 \cite{DF1}. 

%%%%%%%%%%%%%%%%%%   Fig. 1   %%%%%%%%%%%%%%%%%%%%%%%%%
\begin{figure}[H]
\centering
\includegraphics[width=9cm]{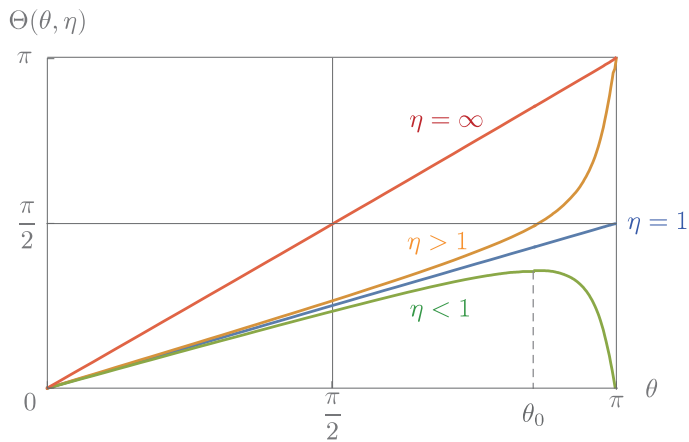}
\caption{The relation between $\theta$ and $\Theta(\theta, \eta)=\theta-\alpha(\theta,\eta)$ parameterized by $\eta$. We have the exact relations $\Theta(\theta, \infty)=\theta$, $\Theta(\theta, 1)=\theta/2$ and $\Theta(\theta, 0)=0$, respectively, for $\eta=\infty$, $\eta=1$ and $\eta=0$. Topologically, $\eta>1$ corresponds to a monopole, $\eta=1$ corresponds to a half-monopole, and $\eta<1$ corresponds to a dipole, respectively \cite{DF1}.  
Note that $\cos \theta_0 =-\eta$  with $\eta<1$, for which $\partial\Theta(\theta, \eta)/\partial\theta=0$. }
\end{figure}
%%%%%%%%%%%%%%%%%%   Fig. 1   %%%%%%%%%%%%%%%%%%%%%%%%%

The Dirac string appears at the singularity of the potential \eqref{new potential}. There exists no singularity at $\theta=0$ since $\Theta(\theta, \eta)\rightarrow 0$ for $\theta\rightarrow 0$.
The singularity does not appear at the origin $B=0$ with any fixed $T$ since $\alpha(\theta,\eta)\rightarrow \theta$ for $B\rightarrow 0$, namely, if one uses $\Theta(\theta, \eta)\rightarrow 0$ for $\eta=\mu BT/\pi\rightarrow 0$ in \eqref{cotangent}. In fact the potential vanishes at $B=0$ for any finite $T$; we have a useful relation in the non-adiabatic domain $\eta=\mu BT/\pi \ll 1$ \cite{DF1},
\begin{eqnarray}
{\cal A}_{\varphi}
&\simeq& \frac{\hbar}{4B}(\mu TB/\pi)^{2}\sin\theta
\end{eqnarray}  
that has no singularity associated with the Dirac string at $\theta=\pi$ near $B=0$ and vanishes at $B=0$.
  Thus the Dirac string can appear only at $\theta=\pi$ and only when $\Theta(\pi, \eta)\neq 0$, namely, $\eta\geq 1$ in Fig.1 or equivalently  
\begin{eqnarray}
 B\geq \frac{\pi}{\mu T}
 \end{eqnarray}
for any fixed finite $T$ \cite{DF1}; the end of the Dirac string is located at $\frac{\pi}{\mu T}$ and $\theta=\pi$.  The total magnetic flux passing through a small circle C  around the Dirac string at the point $B$ and $\theta=\pi$
is given by the potential \eqref{new potential}
\begin{eqnarray}\label{Stokes}
\oint_{C} {\cal A}_{\varphi} B\sin\theta d\varphi =\frac{e_{M}}{2}(1-\cos\Theta(\pi, \eta))
\end{eqnarray}
with $e_{M}=2\pi\hbar$. This flux agrees with the integrated flux outgoing from a sphere with a radius $B$ covering the monopole
due to Stokes' theorem, since no singularity appears except for the Dirac string.  For $\eta>1$, one sees from Fig.1 that the above flux is given by $e_{M}=2\pi\hbar$ and thus Dirac's quantization condition is satisfied in the sense $\exp[-ie_{M}/\hbar]=1$. On the other hand, the flux vanishes for $\eta<1 (i.e., B<\frac{\pi}{\mu T})$ and thus the object changes to a dipole \cite{DF1}.

\subsection{Fixed $T$ configurations}                                                                

We analyze the behavior of the magnetic monopole-like object \eqref{new potential} for fixed $T$ and varying B;
this is close to the description of a monopole in the real space if one identifies $B$ with the radial variable $r$ of the real space.  
The topology and topology change of Berry's phase when regarded as a magnetic monopole defined in the space of $\vec{B}$ is specified by the parameter $\eta$, as is suggested  by a discrete jump of the end point $\lim_{\theta\rightarrow\pi}\Theta(\theta, \eta)$ in Fig.1 \cite{DF1}. 

 Using the exact potential \eqref{new potential}
we have an analogue of the magnetic flux in the {\em parameter space} $\vec{B}=B(\sin\theta\cos\varphi, 
\sin\theta\sin\varphi,\cos\theta )$, 
\begin{eqnarray}\label{magnetic flux}
{\cal B}\equiv \nabla\times {\cal A}
&=&\frac{\hbar}{2}\frac{\frac{\partial\Theta(\theta,\eta)}{\partial\theta}\sin\Theta(\theta,\eta)}{\sin\theta}\frac{1}{B^{2}}{\bf e}_{B} - \frac{\hbar}{2}\frac{\frac{\partial\Theta(\theta,\eta)}{\partial B}\sin\Theta(\theta,\eta)}{B\sin\theta}{\bf e}_{\theta}
\end{eqnarray}
for $\theta\neq \pi$ and $B\neq 0$ with ${\bf e}_{B}=\frac{\vec{B}}{B}$, and  ${\bf e}_{\theta}$ is a unit vector in the direction $\theta$ in the spherical coordinates.
We have
\begin{eqnarray}  
\frac{\partial\Theta(\theta,\eta)}{\partial\theta} = \frac{\eta(\eta+\cos\theta)}{1+\eta^{2}+2\eta\cos\theta},
\end{eqnarray}
by noting $
\frac{\partial\alpha(\theta,\eta)}{\partial\theta}=\frac{1+\eta\cos\theta}{(\eta+\cos\theta)^{2}+\sin^{2}\theta}$ in \eqref{cotangent},
and thus 
$
\frac{\partial\Theta(\theta,\eta)}{\partial\theta}=0$ 
at $\cos\theta_{0}=-\eta$ for $\eta<1$.
The factor in the  second term in \eqref{magnetic flux} is given by recalling $\eta=\mu TB/\pi$, 
\begin{eqnarray}
\frac{\partial\Theta(\theta,\eta)}{\partial B}&=&\frac{\mu T}{\pi}\frac{\partial\Theta(\theta,\eta)}{\partial \eta}\nonumber\\
&=&\frac{\eta}{B}\frac{\sin\theta}{1+\eta^{2}+2\eta\cos\theta}
\end{eqnarray}
using \eqref{cotangent} and \eqref{effective angle}. Thus we have (by setting $e_{M}=2\pi\hbar$)
\begin{eqnarray}\label{magnetic flux2}
{\cal B}&\equiv& \nabla\times {\cal A}\nonumber\\
&=&\frac{e_{M}}{4\pi}\frac{\sin\Theta(\theta,\eta)}{\sin\theta}\frac{\eta}{B^{2}}\frac{1}{1+\eta^{2}+2\eta\cos\theta}[(\eta+\cos\theta){\bf e}_{B} -\sin\theta{\bf e}_{\theta}]
\end{eqnarray}
We also have from \eqref{cotangent},
\begin{eqnarray}\label{alpha parameter}
\cos\alpha=\frac{\eta+\cos\theta}{\sqrt{1+\eta^{2}+2\eta\cos\theta}}, \ \ \  \sin\alpha=\frac{\sin\theta} {\sqrt{1+\eta^{2}+2\eta\cos\theta}},
\end{eqnarray} 
and thus 
\begin{eqnarray}
\sin\Theta(\theta,\eta)&=&\sin(\theta-\alpha) \nonumber\\
&=&\sin\theta\cos\alpha -\cos\theta\sin\alpha\nonumber\\
&=&\frac{\eta\sin\theta}{\sqrt{1+\eta^{2}+2\eta\cos\theta}} ,\end{eqnarray}
and similarly $\cos\Theta(\theta,\eta)
=[1+\eta\cos\theta]/\sqrt{1+\eta^{2}+2\eta\cos\theta}$.

We finally have the azimuthally symmetric magnetic field from \eqref{magnetic flux2}
\begin{eqnarray}\label{magnetic flux3}
{\cal B}
&=&\frac{e_{M}}{4\pi}\frac{\eta^{2}}{B^{2}}\frac{1}{(1+\eta^{2}+2\eta\cos\theta)^{3/2}}[(\eta+\cos\theta){\bf e}_{B} -\sin\theta{\bf e}_{\theta}].
\end{eqnarray}
We note that $B/\eta=\pi/\mu T$ and $\theta=\pi$ define the end  point of the Dirac string in the fixed $T$ picture. The magnetic field
${\cal B}$  is not singular at $\theta=\pi$ for $\eta>1$ which shows that the Dirac string is not observable if it satisfies the Dirac quantization condition. In the adiabatic limit $\eta\rightarrow\infty$ ($\pi/\mu T\rightarrow 0$ with fixed $B$) in \eqref{magnetic flux3}, the outgoing magnetic flux agrees with that of the Dirac monopole
\begin{eqnarray}
 {\cal B}
=\frac{e_{M}}{4\pi}\frac{1}{B^{2}}{\bf e}_{B}
\end{eqnarray}
 located at the origin (level crossing point)  in the parameter space. This is the common magnetic monopole field associated with Berry's phase in the precise adiabatic approximation. At the origin $B=0$ with fixed finite $T$, which corresponds to the nonadiabatic limit $\eta=\mu BT/\pi\rightarrow 0$, the magnetic field \eqref{magnetic flux3} approaches a constant field parallel to the z-axis 
 \begin{eqnarray}
 {\cal B}
=\frac{e_{M}}{4\pi}(\frac{\mu T}{\pi})^{2}[\cos\theta{\bf e}_{B} -\sin\theta{\bf e}_{\theta}].
\end{eqnarray} 
A view of the magnetic flux generated by the monopole-like object \eqref{magnetic flux3} is shown in Fig.2. 

 In passing, we comment on the notational conventions: $\vec{B}(t)$ stands for the externally applied magnetic field to define the original Hamiltonian in \eqref{starting equation} and $\vec{B}$ is used to specify the parameter space to define Berry's phase, and ${\cal B}$ stands for the ``magnetic field'' generated by Berry's phase in the parameter space. The calligraphic symbols ${\cal A},\ {\cal B}$, ${\cal \nabla}$ and the bold ${\bf e}$ stand for vectors without arrows.

%%%%%%%%%%%%%%%%%%   Fig. 2   %%%%%%%%%%%%%%%%%%%%%%%%%
\begin{figure}[H]
\centering
\includegraphics[width=6cm]{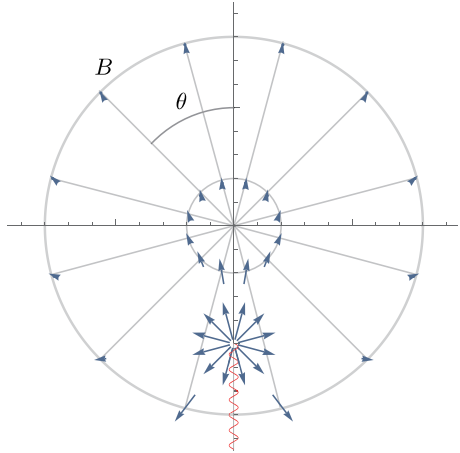}
\caption{Arrows indicating the direction and magnitude of the magnetic flux from the azimuthally symmetric monopole-like object associated with Berry's phase \eqref{magnetic flux3} in the fixed  $T$ picture. Two spheres with radii $B>\pi/\mu T$ (i.e., $\eta>1$) and $B<\pi/\mu T$ (i.e., $ \eta<1$) are shown. The wavy line stands for the Dirac string with the end located at $B=\pi/\mu T$ and $\theta=\pi$ from which the magnetic flux is imported. Only in the ideal adiabatic limit $T\rightarrow\infty$, the end of the Dirac string and the geometrical center of Berry's phase which is located at the origin agree.}
\end{figure}
%%%%%%%%%%%%%%%%%%   Fig. 2   %%%%%%%%%%%%%%%%%%%%%%%%%

\subsection{Lensing of Dirac monopole in Berry's phase}
We show that the monopole associated with Berry's phase is mathematically regarded as a Dirac monopole moving away from the level crossing point of the parameter space driven by the force generated by the nonadiabatic rotating external field with finite period $T=2\pi/\omega<\infty$ in Berry's model.  We consider the configuration in Fig.3. 

%%%%%%%%%%%%%%%%%%   Fig. 3  %%%%%%%%%%%%%%%%%%%%%%%%%
\begin{figure}[H]
\hspace{-4cm}
\includegraphics[width=14cm]{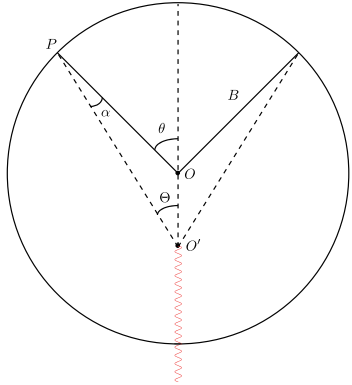}
\caption{A geometric picture in the 3-dimensional parameter space $\vec{B}$ with a sphere centered at $O$ and the radius $B$ by assuming azimuthal symmetry. We suppose that  a genuine azimuthally symmetric Dirac monopole is located at the point $O^{\prime}$ in the parameter space.  The distance between $O$ and $O^{\prime}$ is chosen at 
$\overline{OO^{\prime}} = B/\eta$. The three angles $\theta$, $\alpha$ and $\Theta=\theta-\alpha$ are shown. The observer is located at the point $P$. The wavy line indicates the Dirac string.}
\end{figure}
%%%%%%%%%%%%%%%%%%   Fig. 3  %%%%%%%%%%%%%%%%%%%%%%%%%

We then have 
\begin{eqnarray}
\overline{O^{\prime}P}^{2}
&=&B^{2}+(\frac{B}{\eta})^{2} -2B(\frac{B}{\eta})\cos(\pi-\theta)
\nonumber\\
&=&\frac{B^{2}}{\eta^{2}}[1 + \eta^{2} + 2\eta\cos\theta],
\end{eqnarray}
and the unit vector ${\bf e}$ in the direction of $\vec{O^{\prime}P}$ is 
\begin{eqnarray}
{\bf e}
&=&\cos\alpha {\bf e}_{B} -\sin\alpha {\bf e}_{\theta}
\end{eqnarray}
with $ {\bf e}_{B}=\vec{B}/B$ and $ {\bf e}_{\theta}$ is a unit vector in the direction of $\theta$ in the spherical coordinates.
Then the magnetic flux of Dirac's monopole located at $O^{\prime}$ when observed at the point P is given by 
\begin{eqnarray}\label{Dirac monopole}
{\cal B}^{\prime} &=& \frac{e_{M}}{4\pi}\frac{1}{\overline{O^{\prime}P}^{2}}{\bf e}\nonumber\\
&=&\frac{e_{M}}{4\pi}\frac{\eta^{2}}{B^{2}}\frac{1}{1+\eta^{2}+2\eta\cos\theta}(\cos\alpha {\bf e}_{B} -\sin\alpha {\bf e}_{\theta}).
\end{eqnarray}
Next we fix the parameter $\alpha$. We have $
(B/\eta)^{2}=B^{2} + \overline{O^{\prime}P}^{2} -2B\overline{O^{\prime}P}\cos\alpha$
which gives 
\begin{eqnarray}
\cos\alpha&=&\frac{1}{2B(B/\eta)\sqrt{1 + \eta^{2} + 2\eta\cos\theta}}[B^{2} + (\frac{B}{\eta})^{2}(1 + \eta^{2} + 2\eta\cos\theta)-(\frac{B}{\eta})^{2}]\nonumber\\
&=&\frac{\eta+\cos\theta}{\sqrt{1 + \eta^{2} + 2\eta\cos\theta}}
\end{eqnarray}
and  from the geometrical relation $\frac{B\sin\alpha}{B\sin\theta}=\frac{B/\eta}{(B/\eta)\sqrt{1 + \eta^{2} + 2\eta\cos\theta}}$ ,
\begin{eqnarray}
\sin\alpha=\frac{\sin\theta}{\sqrt{1 + \eta^{2} + 2\eta\cos\theta}}.
\end{eqnarray}
 The parameter $\alpha$ agrees with the parameter in \eqref{alpha parameter}. 
The azimuthally symmetric flux \eqref{Dirac monopole} is thus given by 
\begin{eqnarray}\label{Dirac monopole2}
{\cal B}^{\prime}
&=&\frac{e_{M}}{4\pi}\frac{\eta^{2}}{B^{2}}\frac{1}{(1+\eta^{2}+2\eta\cos\theta)^{3/2}}[(\eta+\cos\theta) {\bf e}_{B} -(\sin\theta){\bf e}_{\theta})]
\end{eqnarray}
which agrees with the flux given by Berry's phase \eqref{magnetic flux3}.

This agreement of two expressions \eqref{magnetic flux3} and \eqref{Dirac monopole2}  shows that the Dirac monopole originally at the level crossing point in the parameter space  formally appears to drift away by the distance $B/\eta=\pi/\mu T$ in the parameter space when the precise adiabaticity  condition $T=\infty$ \cite{Simon} is spoiled by the finite $T$.  
It is interesting  
that two dynamical parameters,  the strength of the external magnetic field and the period in Berry's model, are converted to very different geometrical parameters in Berry's phase, namely, the shape of the monopole and the distance of the deviation of the monopole from the level crossing point. 
The observed magnetic field on the sphere with a radius $B$, which is controlled by the observer, thus changes when one changes the parameter $T$ that determines the end of the Dirac string located at $\pi/\mu T$ in the parameter space.  
This geometrical picture is useful when one draws the precise magnetic flux from the monopole-like object for finite $T$ as in Fig.4 and it is essential when one attempts to understand the motion of a particle in the magnetic field.

%%%%%%%%%%%%%%%%%%   Fig. 4  %%%%%%%%%%%%%%%%%%%%%%%%%
\begin{figure}[H]
\centering
\includegraphics[width=14cm]{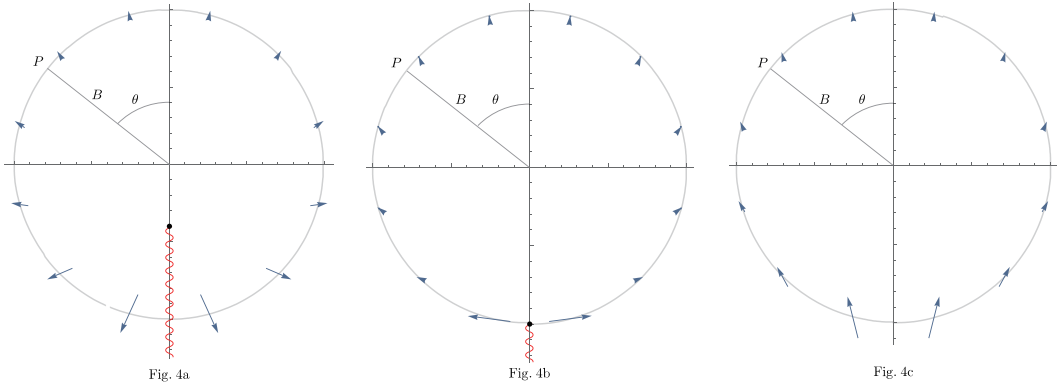}
\caption{Arrows indicating the direction and magnitude of the  magnetic flux observed at the point P with fixed $B$ and $\theta$ when the end of Dirac string at $\pi/\mu T$ is varied from the point $\pi/\mu T<B$ (Fig.4a) to the boundary $\pi/\mu T=B$ (Fig.4b) and then to the point $\pi/\mu T>B$ (Fig.4c), which correspond to the change of the basic parameter $\eta=\mu BT/\pi$ from the adiabatic domain $\eta=2.5>1$ to the boundary  $\eta=1$ and then to the nonadiabatic domain $\eta=0.5<1$, respectively. The wavy line stands for the Dirac string with the end at $B=\pi/\mu T$ and $\theta=\pi$. These figures after a suitable rescaling may also be interpreted as the results with the end of the Dirac string kept fixed at $\pi/\mu T$ and $\theta=\pi$ and varying the distance $B$, starting with a large $B>\pi/\mu T$  (Fig.4a) toward a small $B<\pi/\mu T$  (Fig.4c) in the parameter space, such as two spheres in Fig.2.}
\end{figure}
%%%%%%%%%%%%%%%%%%   Fig. 4  %%%%%%%%%%%%%%%%%%%%%%%%%

In terms of the original physical setting of a magnetic dipole placed in a given rotating magnetic field described by the Hamiltonian \eqref{starting equation}, the cone drawn by the dipole becomes  sharper compared to the cone of the given magnetic field, which subtends the solid angle $\Omega=2\pi (1-\cos\theta)$, when the rotating speed of the external magnetic field becomes larger and the dipole moment is left behind, namely, \cite{Fujikawa1}
\begin{eqnarray}
\psi_{+}^{\dagger}(t)\vec{\sigma}\psi_{+}(t)&=&
w_{+}^{\dagger}(t)\vec{\sigma}w_{+}(t)\nonumber\\
 &=&\left( \sin\Theta\cos\varphi(t), \sin\Theta\sin\varphi(t), \cos\Theta \right)\nonumber\\
&=&-\psi_{-}^{\dagger}(t)\vec{\sigma}\psi_{-}(t)
\end{eqnarray}
that subtends the solid angle 
$\Omega=2\pi (1-\cos\Theta)$ with $\Theta=\theta-\alpha$
; this sharper cone is effectively recognized as the drifting monopole in Berry's phase by an observer located at the point $P$ in Fig.3. 

We note that the agreement of the solid angle drawn by the spinor solution \eqref{exact eigenfuntion} with Berry's phase is known to be generally valid in the two-component spinor. The general orthonormal spinor bases are parameterized as 
\begin{eqnarray}
v_{+}(t)&=&\left(\begin{array}{c}
            \cos\frac{1}{2}\theta(t) e^{-i\varphi(t)}\\
            \sin\frac{1}{2}\theta(t)
            \end{array}\right), \ \ \ 
v_{-}(t)=\left(\begin{array}{c}
            \sin\frac{1}{2}\theta(t)e^{-i\varphi(t)}\\
            -\cos\frac{1}{2}\theta(t)
            \end{array}\right)\nonumber
\end{eqnarray}
that give the spin vector
\begin{eqnarray}
v_{+}^{\dagger}(t)\vec{\sigma}v_{+}(t)
 &=&\left( \sin\theta(t)\cos\varphi(t), \sin\theta(t)\sin\varphi(t), \cos\theta(t) \right)=-v_{-}^{\dagger}(t)\vec{\sigma}v_{-}(t)\nonumber
\end{eqnarray}
subtending the solid angle 
$\tilde{\Omega}_{\pm}=\oint (1\mp\cos\theta(t))d\varphi(t)$ for a closed movement. On the other hand, the ``holonomy'', which is related to Berry's phase, satisfies \cite{Fujikawa1}
\begin{eqnarray}
\oint dt v^{\dagger}_{\pm}(t)i\partial_{t}v_{\pm}(t)=-\frac{1}{2}\oint  (1\mp \cos\theta(t))d\varphi(t) +2\pi =-\frac{1}{2}\tilde{\Omega}_{\pm} +2\pi.\nonumber
\end{eqnarray}
These two quantities thus agree up to the factor $1/2$ and up to trivial phase $2\pi$ in the case of spinor bases. The important fact is that our exact solution of the Schr\"{o}dinger equation \eqref{exact eigenfuntion} has this structure of $v_{+}(t)$ and $v_{-}(t)$ with $\theta(t)=\theta-\alpha(\theta)$. 

One may thus prefer to understand that Fig.3 implies an analogue of the effect of lensing of Dirac's monopole, since the movement of the monopole in the parameter space is a mathematical one.  In the precise adiabatic limit with $T=\infty$ \cite{Simon}, the monopole is located at the level crossing point $O$, but when the effect of nonadiabatic rotation with finite $T< \infty$ is turned on, the image of the monopole is displaced to the point $O^{\prime}$ located at $\pi/\mu T$ by keeping the  topology and strength of the point-like monopole intact. In this picture, it is important that the topological monopole itself is not resolved in the nonadiabatic domain but it  disappears from  observer's view located at the point P for fixed $B$ when $\pi/\mu T = B/\eta \rightarrow {\rm large} $ with fixed $B$ (i.e., $\eta\rightarrow$ small). In the middle, the formal topology change takes place when $\pi/\mu T$ touches the sphere with the fixed radius $B$ (i.e., $\eta=1$). Even in the picture of lensing, the ``magnetic flux'' generated by Berry's phase  measured at the point in the parameter space specified by $(B,\theta)$ is the real flux.
It will be interesting to examine the possible experimental implications of these aspects of Berry's phase, which is expressed  by the magnetic field \eqref{magnetic flux3}, in a wider area of physics. 

As for the smooth transition from a monopole to a dipole, it appears in the process of shrinking of the sphere with a radius $B$ covering the end of the Dirac string located at $\pi/\mu T$ to a smaller sphere for which $B<\pi/\mu T$ as in Fig.2. When the sphere touches the end of the Dirac string (at $\eta=1$) in the middle, one encounters a  half monopole with the outgoing flux which is half of the full monopole $e_{M}/2=\pi\hbar$. See Stokes' theorem \eqref{Stokes} with $\Theta(\pi, \eta=1)=\pi/2$ in Fig.1. At this specific point, the Dirac string becomes observable \cite{DF1}, corresponding to the Aharonov-Bohm effect \cite{Aharonov} of the electron in the magnetic flux generated by the superconducting Cooper pair \cite{Tonomura}.  
It is then natural to attach the end of the Dirac string to an infinitesimally small opening on the sphere forming a closed sphere and thus leading to the vanishing net outgoing flux, which corresponds to a dipole. The idea of the half monopole at $\eta=1$ is interesting, but it is natural to incorporate it as a part of a dipole. The monopole-like object \eqref{new potential} is always  a dipole if one counts the Dirac string as in Fig.2  and Stokes' theorem \eqref{Stokes} always holds.  In this sense, no real topology change takes place for the movement of $B$, from large $B$ to small $B$, except for the fact that the unobservable Dirac string becomes observable at $B=\pi/\mu T$ and triggers the topology change from a monopole to a dipole.

\section{Discussion and conclusion}

The topology or singularity in Berry's phase arises from the well-known adiabatic theorem \cite{Born, Kato}, namely, no level crossing takes place in the precise adiabatic limit $T\rightarrow \infty$. This theorem implies the appearance of some kind of obstruction or barrier to the level crossing in the precise adiabatic limit; the appearance of Dirac's monopole singularity in the adiabatic limit  may be regarded as a manifestation of this obstruction or barrier in the parameter space.  Off the precise adiabatic limit with finite $T$, which is physically relevant  for the applications of Berry's phase as was noted by Berry \cite{Berry}, no more  obstruction to the level crossing appears. This is a basis of our expectation of the topology change in Berry's phase in the nonadiabatic domain.

The topology change in Berry's phase in the exactly solvable model has been analyzed in detail including the appearance of a half-monopole in \cite{DF1}. The analysis is essentially based on Fig.1 that is a result of solving 
the relation \eqref{cotangent}, which is in turn a result of the Schr\"{o}dinger equation \eqref{starting equation}. Because of this complicated logical procedure, the exact ``magnetic field'' generated by Berry's phase was not very transparent.  In the present paper, we remedied this short coming in \cite{DF1} by giving a more explicit representation of the magnetic field. In this attempt, we recognized that the magnetic field is in fact given by a very simple geometrical picture in Fig.3.  We thus encountered an interesting mathematical description of the topology change in Berry's phase in terms of a geometrical movement of Dirac's monopole caused by the nonadiabatic variations of parameters in Berry's model. It is remarkable that the monopole remains in tact without being resolved even in the nonadiabatic domain.
We analyzed the monopole-like object and its topological property in Berry's phase by treating $\vec{B}$ as a given classical parameter.  If one adds other physical considerations, there appear some conditions on the parameters of the exactly solvable model \eqref{starting equation}. For example, the two levels in \eqref{energy eigenvalue} cross at $\alpha(\theta;\eta)=\pi/2$, which is related to the topology change from a monopole to a dipole in an intricate way if one remembers $\alpha(\theta;\eta)=\theta-\Theta$ and Fig.1.

Traditionally, we are accustomed to understanding the topology change in terms of the winding and unwinding of some topological obstruction. The present geometrical description of topology change in terms of the moving monopole is a hitherto unknown mechanism. This new mechanism  partly arises from the fact that Berry's phase is not a simple monopole but rather a complex of the monopole and the level crossing point located at the origin of coordinates.  If one instead understands Berry's phase as a simple monopole, one will find a novel class of monopoles \cite{DF2, Mavromatos}.

A notable application of Berry's phase in momentum space, which is defined by the effective Hamiltonian by replacing 
$\vec{B}(t)\rightarrow \vec{p}(t)$ in the original model of Berry 
\begin{eqnarray}\label{condensed matter}
\hat{H}=-\mu \vec{\sigma}\cdot \vec{p}(t),
\end{eqnarray}
is known in the analyses of the anomalous Hall effect \cite{Jungwirth} and the spin Hall effect \cite{Hirsch}.  
This effective Hamiltonian of the two-level crossing for the generic $\vec{p}(t)$ (Bloch momentum) has been analyzed in detail in \cite{DF1}, and it has been shown that Berry's phase for \eqref{condensed matter} is determined by the time derivative of the azimuthal angle $\dot{\varphi}(t)$ in both adiabatic (monopole) and nonadiabatic (dipole) limits, and thus our parameterization \eqref{solvable model} describes an essential aspect of the topology of Berry's phase. To be more precise, Berry's phase becomes trivial, namely, either $0$ or $2\pi$, in the model \eqref{condensed matter} for the nonadiabatic limit \cite{DF1}
\begin{eqnarray}\label{bound in generic model}
(\mu |\vec{p}|)T/\hbar \ll 1
\end{eqnarray}
which corresponds to $\eta\ll 1$ in terms of the parameter in \eqref{parameter2}.  This estimate is consistent with the analysis of the exactly solvable model for $\eta\rightarrow 0$ for which $\Theta\rightarrow 0$ in Fig. 1, and thus 
$\Omega_{\pm}/\hbar \rightarrow 0\ {\rm or}\ 2\pi$ in \eqref{solid-angle}.
Our present analysis implies that one may be able to observe experimentally the effective movement of the monopole in momentum space, as is represented by the magnetic field in \eqref{magnetic flux3} (by replacing $B\rightarrow |\vec{p}|$), at away from the precise adiabaticity in the model \eqref{condensed matter}. 
Also, it will be interesting to examine the implications of the present analysis on the very basic issue if Berry's phase associated with \eqref{condensed matter} deforms the principle of quantum mechanics by giving rise to anomalous canonical commutators \cite{DF3}.

In conclusion, the analysis of an exactly solvable model has  revealed that the topology change in Berry's phase is mathematically visualized as the geometrical movement or the lensing of Dirac's monopole in the parameter space. This will help better understand both Berry's phase and  Dirac's monopole. 
\\

The present work is supported in part by JSPS KAKENHI (Grant No.18K03633).


\begin{thebibliography}{99}
\bibitem{Higgins}
H. Longuet-Higgins, Proc. Roy. Soc. A{\bf 344}, 147 (1975).
\bibitem{Berry}
M. V.  Berry, Proc. R. Soc. Ser. A{\bf 392}, 45 (1984).
\bibitem{Simon}
B. Simon, Phys. Rev. Lett. {\bf 51} (1983) 2167.
\bibitem{Dirac}
P.A.M. Dirac, Proc. Roy. Soc. London {\bf 133}, 60 (1931).
\bibitem{Fujikawa1}
K. Fujikawa, Int. J. Mod. Phys. A{\bf 21} (2006) 5333;
K. Fujikawa, Ann. of Phys. {\bf 322}, 1500 (2007).
Earlier works on the basic aspects of Berry's phase are quoted in these references.
\bibitem{DF1}
S. Deguchi and K. Fujikawa, Phys. Rev. D{\bf 100}, 025002 (2019).
\bibitem{Wu-Yang}
T. T.  Wu and C. N. Yang, Phys. Rev. D{\bf 12}, 3845 (1975).
\bibitem{Aharonov}
 Y. Aharonov and D. Bohm, Phys. Rev. {\bf 115}, 485 (1959).

\bibitem{Tonomura}
A. Tonomura, N. Osakabe, T. Matsuda, T. Kawasaki, J. Endo, S. Yano, and H. Yamada,  Phys. Rev. Lett. {\bf  56}, 792    (1986).
\bibitem{DF2}
S. Deguchi and K. Fujikawa, Phys. Lett. B{\bf 802}, 135210 (2020).
\bibitem{Mavromatos}
A recent review of the magnetic monopole is found in N. E. Mavromatos and V. A. Mitsou, ``Magnetic monopoles revisited: Models and searches at colliders and in the Cosmos'',
Int. J. Mod. Phys. A {\bf 35} (2020) 2030012. 
\bibitem{Jungwirth}
T. Jungwirth, Q. Niu, A.H. MacDonald, Phys. Rev. Lett. {\bf 88} (2002) 207208.\\
Z. Fang, et al., Science {\bf 302} (2003) 92, and references therein.
\bibitem{Hirsch}
J.E. Hirsch, Phys. Rev. Lett. {\bf 83} (1999) 1834;
S.-F. Zhang, Phys. Rev. Lett. {\bf 85} (2000) 393;
S. Murakami, N. Nagaosa, S.-C. Zhang, Science {\bf 301} (2003) 1348.
%\bibitem{Niu}
%D. Xiao, J. Shi, Q. Niu, Phys. Rev. Lett. {\bf 95} (2005) 137204.
\bibitem{Born}
M. Born and V. Fock, Zeitschrift f. Phys. {\bf 51} (1928) 165.
\bibitem{Kato}
T. Kato, J. Phys. Soc. Jpn. {\bf 5}, 435 (1950).
\bibitem{DF3}
S. Deguchi and K. Fujikawa, Ann. of Phys. {\bf 416} (2020) 168160.

\end{thebibliography}
\end{document}